# Clifford Theory: A Geometrical Interpretation of Multivectorial Apparent Power

M. Castilla, J. C. Bravo, M. Ordóñez, and J. C. Montaño, *Senior Member IEEE*

*Abstract*— In this paper, a generalization of the concept of electrical power for periodic current and voltage waveforms based on a new generalized complex geometric algebra (GCGA), is proposed. This powerful tool permits, in *n*-sinusoidal/ nonlinear situations, representing and calculating the voltage, current, and apparent power in a single-port electrical network in terms of *multivectors*. The new expressions result in a novel representation of the apparent power, similar to the Steinmetz's phasor model, based on complex numbers, but limited to the purely sinusoidal case. The multivectorial approach presented is based on the frequency domain decomposition of the apparent power into three components: the real part and the imaginary part of the complex-scalar associated to active and reactive power respectively, and distortion power, associated to the complex-bivector.
A geometrical interpretation of the multivectorial components of apparent power is discussed. Numerical examples illustrate the clear advantages of the suggested approach.

*Index Terms*— Clifford algebra, vector-phasor, multivectorial apparent power, harmonics.

## I. INTRODUCTION

THE application of Geometric Algebra (GA) [1-2] to electromagnetic theory, in which circuit analysis is a natural consequence, has a very short history. On the contrary, the classical power theory has been widely analyzed, but some fundamental concepts are still unsolved. In sinusoidal conditions, power equation is described by complex algebra and decomposed into apparent, active and reactive powers. For
the *n*-sinusoidal case, research on power definitions [3]-[14] has been carried out with very different objectives as mathematical meaning, physical meaning, power factor improvement, distortionless conditions, etc. Several recent papers have dealt with the definition and compensation of nonactive power [15]-[17], but the old contributions of Budeanu in frequency domain [3] and Fryze [4] in time domain remain essential. The large number of papers published motivated by these two classic theories suggest that the work has not been finished. Unfortunately, many contributions do not consider the multivectorial character of the apparent power components. Other authors [18],[19],[28], have proposed new power equations based in vector spaces representation. In particular [19] is concerned with a representation of power equation in the mathematical framework of Geometric Algebra. Nevertheless, this last reference only uses the information of impedance angles without including voltage phase angles. Therefore, it has been concluded that the typical nonlinear behaviour of the distribution systems require, for its complete analysis, a new mathematical structure that can guarantee the multivectorial character of different components.

In this sense, our work considers a new representation of power theory deduced from a generalized Geometric Algebra. It is based on a decomposition of apparent power into multivectorial components in the frequency domain. The apparent power multivector is derived in terms of the voltage and current vector-phasors, and contains all power information (magnitude, direction, and sense). Previously, the voltage and current waveforms have been transformed into the frequency domain via Clifford Fourier transform [20]. In particular, the phase shift of voltage vector-phasor is considered.

The geometrical interpretation of power components is quite effective for clarifying their nature from a mathematical and physical viewpoint. Particularly, the new mathematical framework GCGA is aimed at yielding the following general contributions:

• The introduction of a new mathematical structure for a clear definition of the concept of vector-phasor applied to algebraic and geometric representation of voltage and current signals in frequency domain analysis.

• A new frequency-domain analysis of the multivector quantities associated with complex geometric algebra, in order to explain the concept of instantaneous power.

• Geometrical interpretation of Power Theory and the new features of electrical power decomposition into active, reactive, and distortion powers.

• Definition of the distortion power in multivectorial form.

Manuscript received June 19, 2007. This work was supported by the Ministry of Education and Science as part of a research through project DPI-2006-17467-CO2-01.

M. Castilla and J. C. Bravo are with the Electrical Engineering Department of the University of Sevilla, Escuela Univ. Politécnica, C/Virgen de África 7, (41011) - Sevilla, (Spain). (phone:+34 954 55 28 47; fax: +34 954 55 16 88; e-mail: castilla@.us.es; carlos_bravo@us.es).
M. Ordóñez is with the Department of Applied Mathematics II. Escuela, Univ. Politécnica, C/Virgen de África, 7, (41011) - Sevilla, (Spain). (e-mail:mordonez@us..es
J. C. Montaño is with the Spanish Research Council (CSIC), Reina Mercedes Campus, POB 1052, 41080-Sevilla, Spain. (email:montano@irnas.csic.es).

- Operational facility for the analysis of the linear and nonlinear, and time-variant and time-invariant networks.
- Reversibility frequency domain- time domain via Clifford Fourier transform [20].
- Finally, the possibility of extending to power multivector concept in poly-phase systems.

In order to help the reader, the complete mathematical foundations have been shown in the Appendix.

## II. CLIFFORD SPACE-VECTOR THEORY: GENERALIZED COMPLEX GEOMETRIC ALGEBRA ($CCl_n$)

### A. Preliminaries: classic geometric product $(g)$

Geometric algebras can be defined simply by specifying appropriate rules for multiplying vectors. Thus, let $V^n$ an n-dimensional linear space over the real numbers. The geometric product of vectors $(\mathbf{a}) g (\mathbf{b})$ or $\mathbf{ab}$ if $\mathbf{a},\mathbf{b} \in V^n$ can be decomposed into *a symmetric* inner product

$$\mathbf{a} \cdot \mathbf{b} = \frac{1}{2}(\mathbf{ab} + \mathbf{ba}) \quad (1)$$

and an *antisymmetric* outer product

$$\mathbf{a} \wedge \mathbf{b} = \frac{1}{2}(\mathbf{ab} - \mathbf{ba}) \quad (2)$$

Therefore, $\mathbf{ab}$ has the *canonical decomposition*

$$\mathbf{ab} = \mathbf{a} \cdot \mathbf{b} + \mathbf{a} \wedge \mathbf{b} \quad (3)$$

The inner product $\mathbf{a} \cdot \mathbf{b}$ is a scalar and the outer product $\mathbf{a} \wedge \mathbf{b}$ is called bivector (or 2-vector). Geometrically, it represents a directed plane segment, in much the same way as a vector represents a directed line segment, fig.1. We can regard $\mathbf{a} \wedge \mathbf{b}$ as a directed area with a magnitude $|\mathbf{a} \wedge \mathbf{b}|$ equal to the usual scalar area of the parallelogram in fig.1, with the direction of the plane in which the parallelogram lies, and with sense which can be assigned to the parallelogram in the plane. Then, just as a vector $\mathbf{a}$ represents (or is represented by) a directed line segment and a bivector $\mathbf{a} \wedge \mathbf{b}$ represents a directed plane segment, the trivector (3-vector) $\mathbf{a} \wedge \mathbf{b} \wedge \mathbf{c}$ represents a directed space segment (the parallelepiped with edges a, b, c)

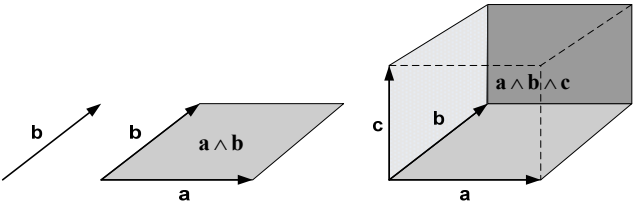

Fig.1 Vector, bivector and trivector representation.

### B. Generalized Complex Geometric Algebra: new geometric product $(\hat{g})$

Let us introduce *vector- phasors* (complex-vectors) in order to analyze circuit power theory in nonsinusoidal conditions. To define these new *phasors*, we start from an *n*-dimensional linear space $V^n$, of elements that are termed vectors. If $\{\sigma_1, \sigma_2, \sigma_3, ... \sigma_n\}$ is an orthonormal basis of $V^n$, (n is equal to the number of harmonic components in periodic non sinusoidal signals), the unit element of this algebra is denoted by $\sigma_0$. The vector basis for the Clifford algebra $\{Cl_n\}$ is generated by

$$\left\{ \underbrace{1}_{scalar}, \underbrace{\sigma_k}_{\substack{vectors \\ (k:1,...,n)}}, \underbrace{\sigma_k \wedge \sigma_h}_{\substack{bivectors \\ (k,h:1,...,n; k \neq h)}}, ..., \underbrace{\sigma_1 \wedge \sigma_2 \wedge \sigma_3 .. \wedge \sigma_n}_{pseudoscalar} \right\} \quad (4)$$

where $\wedge$ denotes the outer product and $\sigma_k \wedge \sigma_h = \sigma_k \sigma_h = \sigma_{kh}$. Each coefficient of a basic vector $\sigma_j$ replaces one of the orthonormal functions in the Fourier decomposition. The elements in this geometric algebra are termed multivectors [1-2]. But the electrical quantities voltage and current have not easy interpretation in classic Clifford Algebra. For this reason we will define a new geometric algebra — a generalization of the classic Clifford Algebra [2] — which we have termed "*Generalized Complex Geometric Algebra*" (GCGA). Then, let $C$ the complex vector space and $Cl_n$ the Clifford algebra in *n*-dimensional real space $V^n$ and the following structure is defined

$$\{CCl_n, \hat{g}\} \quad (5)$$

which coefficients $\overline{z}_{12...k} \in C$, the basis $\sigma_{12...k} \in Cl_n$ and $\hat{g} = (\Re \circ g)$ can be seen in (A1) and (B2-B5). It is trivial that $CCl_n$ is a vector space over $\mathbf{R}$. The generic element $\tilde{Z}_p = \overline{z}_p \sigma_p \in CCl_n$ is a *p-th complex-vector* and it can be represented by $|\overline{z}_p| e^{j\alpha_p} \sigma_p$, $\alpha_p$ is the phase angle of $\overline{z}_p$ and $\sigma_p$ is a basis vector. In this sense, the generic element $\tilde{Z}_{pq} = \overline{z}_{pq} \sigma_{pq} \in CCl_n$ is a *pq-th complex-bivector*, and it can be represented by $|\overline{z}_{pq}| e^{j\alpha_{pq}} \sigma_{pq}$, $\alpha_{pq}$ is the phase angle of $\overline{z}_{pq}$ and $\sigma_{pq}$ is a basis bivector.

The objective is that, elements such as $\tilde{Z}_p = \overline{z}_p \sigma_p = |\overline{z}_p| e^{j\alpha_p} \sigma_p$ will be used to represent voltage and current harmonic vector-phasors and elements of the form $\tilde{Z}_{pq} = \overline{z}_{pq} \sigma_{pq} = |\overline{z}_{pq}| e^{j\alpha_{pq}} \sigma_{pq}$ will be used to analyze power components. A more complete information can be seen in [1-2-21-23] and [Appendix A, B]

## III. Multivectorial Apparent Power

### A. Multivectorial representation of periodic signals.

Suppose that a nonsinusoidal voltage

$$u(t) = \sqrt{2} \sum_{p \in L \cup N} U_p \sin(p\omega t + \alpha_p) \qquad (6)$$

is applied to a nonlinear load, fig. 2, where $p$ is the harmonic order of $u(t)$. The resulting current has an instantaneous value given by

$$i(t) = \sqrt{2} \sum_{q \in N \cup M} I_q \sin(q\omega t + \beta_q) \qquad (7)$$

where $q$ is the harmonic order of $i(t)$. It is assumed that a group of voltage harmonics $N$ exist that have corresponding current harmonics of the same frequencies, that components $L$ of the supply voltage exist without corresponding current, and that components $M$ of current exist without corresponding voltages. In linear operation, $\beta_q = \alpha_q - \varphi_q$, $\varphi_q$ is the impedance phase angle and $L = \{\phi\}, M = \{\phi\}$. The capital $U_p$ and $I_q$ represent rms values of $u_p(t)$ and $i_q(t)$.

In the $\{CCl_n, \hat{g}\}$ structure spanned by an orthonormal basis $\{\sigma_1, \sigma_2, \sigma_3, ... \sigma_n\}$, the associated $p$-th harmonic voltage and $q$-th harmonic current can be represented by the vector-phasors:

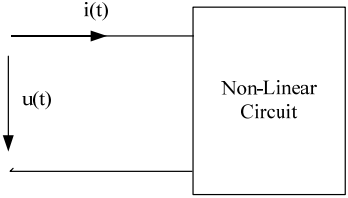

Fig. 2. Generic nonlinear circuit

$$\tilde{U}_p = |\tilde{U}_p| e^{j\alpha_p} \sigma_p = \overline{U}_p \sigma_p$$

and

$$\tilde{I}_q = |\tilde{I}_q| e^{j\beta_q} \sigma_q = \overline{I}_q \sigma_q \qquad (8)$$

where $|\tilde{U}_p| = U_p$, $|\tilde{I}_q| = I_q$. Then $\tilde{U} = \sum_{p \in L \cup N} \tilde{U}_p$, $\tilde{I} = \sum_{q \in N \cup M} \tilde{I}_q$.

Due to the orthonormal character of the Clifford basis, the magnitude of the vector-phasors coincides with the rms value or norm (D1) of $u(t)$ and $i(t)$ signals. The rms values are given by

$$|\tilde{U}|^2 = \sum_{p \in L \cup N} |\tilde{U}_p|^2 \text{ and } |\tilde{I}|^2 = \sum_{q \in N \cup M} |\tilde{I}_q|^2$$

### B. Apparent power multivector

According to (B2-B6), the *apparent power* at the nonlinear load, fig. 2, can be obtained as a multivector $\tilde{S}$ in $V^n$ generated by the geometric product "$\hat{g}$" of the voltage and conjugate current vector-phasors

$$\tilde{S} = \sum_{\substack{p \in N \cup L \\ q \in N \cup M}} \tilde{U}_p \hat{g} \tilde{I}_q^* = \left( \sum_{p=q} U_p I_p \cos\varphi_p + j \sum_{p=q} U_p I_p \sin\varphi_p \right) \sigma_0 +$$
$$+ \sum_{\substack{p<q \\ p,q \in N}} \left\{ \left( U_p I_q e^{j\varphi_q} - U_q I_p e^{j\varphi_p} \right) e^{j(\alpha_p - \alpha_q)} \right\} \sigma_{pq} + \qquad (9)$$
$$+ \sum_{\substack{p \in L \cup N, q \in M \\ p \in L, q \in N}} U_p I_q e^{j(\alpha_p - \beta_q)} \sigma_{pq} = \tilde{P} + j\tilde{Q} + \tilde{D}$$

which consist of a complex-scalar and a complex-bivector. In eqn. (9), "$\hat{g}$" is the new *"generalized complex geometric product"* (B2), and $(*)$ is the standard *"complex conjugate"* operation (C2). Clearly, $|\tilde{P}| = \sum_{p \in N} U_p I_p \cos\varphi_p$ is the active power or average value of the instantaneous power in the time domain. $|\tilde{Q}| = \sum_{p \in N} U_p I_p \cos\varphi_p$ is the called reactive power and is not a real physical quantity. It is merely the geometric complement of active component. Note from eqn. (9) that $(P + jQ)\sigma_0$ is the complex-scalar. The complex-bivector associated to multivectorial distortion power, is given by

$$\tilde{D} = \sum_{\substack{p<q \\ p,q \in N}} \left\{ \left( U_p I_q e^{j\varphi_q} - U_q I_p e^{j\varphi_p} \right) e^{j(\alpha_p - \alpha_q)} \right\} \sigma_{pq} +$$
$$+ \sum_{\substack{p \in L \cup N, q \in M \\ p \in L, q \in N}} U_p I_q e^{j(\alpha_p - \beta_q)} \sigma_{pq} = \tilde{D}_{Lin} + \tilde{D}_{Nonlin} \qquad (10)$$

This complex-bivector $\tilde{D}$ is an entirely fictitious component and non physical variable. The components $\tilde{Q}$ and $\tilde{D}$ have a non independent physical nature and they constitute the nonactive power.

Note that, consistent with (G1), the squared value $|\tilde{S}|^2$ in eqn. (9), may be represented as

$$|\tilde{S}|^2 = |\tilde{U} \hat{g} \tilde{I}^*|^2 = |\tilde{U}|^2 |\tilde{I}|^2 \qquad (11)$$

and

$$|\tilde{S}|^2 = P^2 + Q^2 + D^2 \qquad (12)$$

This expression is identical to the classic squared value of the apparent power.

In linear operation and $\alpha_p = \alpha_q$, eqn.(9) becomes simplified and is now given by

$$\tilde{S} = \sum_{\substack{p \in N \\ q \in N}} \tilde{U}_p g \tilde{I}_q^* = \left( \sum_{p=q} U_p I_p \cos\varphi_p + j \sum_{p=q} U_p I_p \sin\varphi_p \right) \sigma_0 +$$
$$+ \sum_{\substack{p<q \\ p,q \in N}} \left( U_p I_q e^{j\varphi_q} - U_q I_p e^{j\varphi_p} \right) \sigma_{pq} = \tilde{P} + j\tilde{Q} + \tilde{D}_{Lin} \qquad (13)$$

From (B6), note that now "g" is the classic geometric product (3).

The *multivectorial apparent power* $\tilde{S}$, can be given not only as in (9), but also in the form

$$\tilde{S} = \tilde{S}_{Linear} + \tilde{S}_{Nonlinear} = \tilde{P} + \tilde{S}_{Nonactive}, \quad (14)$$

where

$$\tilde{S}_{Linear} = \tilde{P} + j\tilde{Q} + \tilde{D}_{Lin} \quad (15)$$

$$\tilde{S}_{Nonlinear} = \tilde{D}_{Nonlin} = \sum_{\substack{p \in L \cup N, q \in M \\ p \in L, q \in N}} \tilde{D}_{pq} \quad (16)$$

$$\tilde{S}_{Nonactive} = j\tilde{Q} + \tilde{D}_{Lin} + \tilde{D}_{Nonlin} \quad (17)$$

In sinusoidal case, eqns. (8) can be expressed by

$$\tilde{U} = |\tilde{U}_1| e^{j\alpha_1} \sigma_1 = \bar{U}_1 \sigma_1$$

and (18)

$$\tilde{I} = |\tilde{I}_1| e^{j\beta_1} \sigma_1 = \bar{I}_1 \sigma_1$$

where $\tilde{U}_1$ and $\tilde{I}_1$ are now the Steinmetz *classic phasors*. The complex apparent power is defined by

$$\tilde{S} = \tilde{U} \hat{g} \tilde{I}^* = (P + jQ)\sigma_0 \quad (19)$$

where

$$P = U_1 I_1 cos\varphi_1 \quad (20)$$

and

$$Q = U_1 I_1 sin\varphi_1 \quad (21)$$

are *active* and *reactive* powers respectively.

From the viewpoint of the power factor improvement, the suggested decomposition, eqn. (9), can be particularly useful. Thus, a multivectorial *relative quality index (RQI)* is defined by

$$\tilde{\delta} = \frac{\tilde{S}}{\tilde{P}} = 1 + j\frac{\tilde{Q}}{\tilde{P}} + \frac{\tilde{D}}{\tilde{P}} \quad (22)$$

and the *power factor (PF)* can be written as

$$PF = \frac{1}{|\tilde{\delta}|} \quad (23)$$

where

$$|\tilde{\delta}| = \sqrt{1 + \frac{|\tilde{Q}|^2}{|\tilde{P}|^2} + \frac{|\tilde{D}|^2}{|\tilde{P}|^2}} \quad (24)$$

Eqn. (24) shows that on this decomposition, all the components with its direction and sense are accessible in order to improve the power factor.

The suggested apparent power multivector, eqn. (9), is very important and represents a new concept of apparent power. The eqn. (11) is the squared value of $\tilde{S}$, for linear and nonlinear networks under nonsinusoidal conditions. This value $|\tilde{S}|$, is a consequence only of the multivector $\tilde{S}$ and is one of this paper's main contribution. In particular, $|\tilde{S}|^2$ is the sum of the squared values of the components of $\tilde{S}$. It should be noted that whereas $|\tilde{S}|$ is a simple value, the multivector $\tilde{S}$ has magnitude, direction, and sense.

## IV. COMPUTATIONAL ADVANTAGES

Standard mathematical software is not well suited for working with multivectors. In particular, the standard cross product does not work in spaces with dimension greater than 3. Therefore, its use is difficult for analyzing nonsinusoidal circuits, since it is actually necessary to carry out calculations for this generalized multivectors product. The MATHCAD program is very suitable, however, due to its operating procedure symbolic matrix algebra. Using the symbolic method, apparent power in multivectorial form, (9), can be calculated easily if we define the complex coefficients matrix **H**.

$$\mathbf{H} = \begin{Bmatrix} \bar{U}_p \cdot \bar{I}_q^* \cdot e^{-2j(\alpha_p - \alpha_q)} & p > q \; \forall p,q \in N \\ \bar{U}_p \cdot \bar{I}_q^* & otherwise \end{Bmatrix} \quad (25)$$

Thus, $\tilde{S} = \tilde{U} \hat{g} \tilde{I}^* = (\sigma_1, \sigma_2, \cdots, \sigma_n) \mathbf{H} \begin{pmatrix} \sigma_1 \\ \sigma_2 \\ \vdots \\ \sigma_m \end{pmatrix}$. More easily, **H** is given by

$$\mathbf{H} = \begin{pmatrix} \bar{U}_1\bar{I}_1^* & \bar{U}_1\bar{I}_2^* & \cdots & \bar{U}_1\bar{I}_n^* & \bar{U}_1\bar{I}_m^* & 0 \\ \bar{U}_2\bar{I}_1^* e^{-2j(\alpha_2-\alpha_1)} & \bar{U}_2\bar{I}_2^* & \cdots & \bar{U}_2\bar{I}_n^* & \bar{U}_2\bar{I}_m^* & 0 \\ \vdots & \vdots & \ddots & \vdots & \vdots & \vdots \\ \bar{U}_n\bar{I}_1^* e^{-2j(\alpha_n-\alpha_1)} & \bar{U}_n\bar{I}_2^* e^{-2j(\alpha_n-\alpha_2)} & \cdots & \bar{U}_n\bar{I}_n^* & \bar{U}_n\bar{I}_m^* & \vdots \\ 0 & 0 & \cdots & \cdots & 0 & 0 \\ \bar{U}_\ell\bar{I}_1^* & \bar{U}_\ell\bar{I}_2^* & \cdots & \cdots & \bar{U}_\ell\bar{I}_m^* & 0 \end{pmatrix} \quad (26)$$

where the diagonal includes all elements of the inner product and the trace of **H** returns the value of $P + jQ$ directly. On the other hand, the cross elements are used to characterize straightforwardly the new outer product and, hence, the different components of distortion: $\bar{D}_{pq} = \bar{H}_{pq} - \bar{H}_{qp}$.

Note that linear elements conform a submatrix with non zero elements that are affected by phase correction for p>q. Moreover, a nonlinear element introduces a null row (or column) when a voltage (or current) harmonic is not present. In this case, it is not necessary to correct the phase angles. The apparent complexity of the mathematical structure $\{CCl_n, \hat{g}\}$ is explained by the operative facility from computational advantages and intuitive geometrical interpretation of the power theory.

According to (B6), if $\alpha_p = \alpha_q$ for all linear elements, a simplified expression for $\tilde{S}$ multivector and **H** matrix are

given by

$$\tilde{S} = \tilde{U} \, g \, \tilde{I}^*$$

and

$$\mathbf{H} = \{\bar{U}_p \cdot \bar{I}_q^* \quad \forall p,q\}$$

(27)

## V. Geometrical interpretation

In the $(CCl_n, \hat{g})$ algebraic structure, voltage and current vector-phasors (8) must be considered *complex-vectors*, and the multivectorial apparent power (9) is a *complex-multivector* which components can be associated to *active, reactive*, and *distortion powers*. The active and reactive powers result in real and imaginary part respectively within the $\sigma_0$-plane in which all products of the same frequency are contained ($p = q$). The distortion power is a sum of complex bivectors in different $\sigma_{pq}$-planes involving cross-frequency products ($p \neq q$).

For a pair of generic indexes, $p = q$, (fig. 3), we represented the position of the bivectors involving like-frequency geometric products in the $\sigma_0$-plane. The sum of these products is a *complex-scalar* $\tilde{A}_0$, and their components are the active power $P$, and the reactive power $Q$.

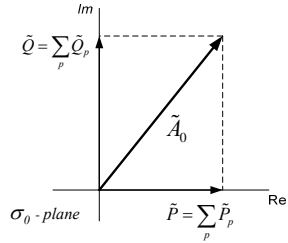

Fig. 3. $\tilde{A}_0$ decomposition into active power $P$ and reactive power $Q$.

If $p \neq q$ and $p, q \in N$, bivectors from cross-frequency geometric products are represented in Fig.4. The difference of these products is a $\tilde{D}_{pq}$ complex-bivector into the $\sigma_{pq}$-plane and is associated to linear distortion power.

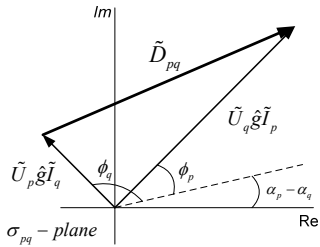

Fig. 4. Representation of the linear multivector distortion: $\tilde{D}_{pq}$

Additionally, if $p \neq q$, and $p \in L \cup N, q \in M$ and $p \in L, q \in N$,

bivectors involving cross-frequency geometric products are associated to non-linear distortion power.

It can be seen from figs. 3 and 4, that the geometric interpretation of eqn. (9) (Budeanu's multivectorial power equation) is based on the association of complex planes $(\mathbf{C})$ to each multivectorial element of Clifford basis $Cl_n$.

## VI. Numerical examples

In this section, two numerical examples are developed. Units of physical quantities are the standard ones of the MKSA system, and are thus omitted everywhere.

### A. Example 1.

A periodical nonsinusoidal voltage with instantaneous value given by

$$u(t)_A = \sqrt{2} \begin{bmatrix} 200\sin\omega t + 200\sin(2\omega t - 30°) \\ + 100\sin(4\omega t + 30°) \end{bmatrix}$$

(28)

is applied to a nonlinear load. The resulting current has an instantaneous value given by

$$i(t)_A = \sqrt{2} \begin{bmatrix} 20\sin(\omega t + 30°) + 10\sin(2\omega t - 60°) \\ + 10\sin(3\omega t + 60°) \end{bmatrix}$$

(29)

Corresponding harmonic vector-phasor expressions of the voltage and conjugated current are given by

$$\tilde{U}_A = 200\, e^{j0}\sigma_1 + 200\, e^{-j30}\sigma_2 + 100 e^{j30}\sigma_4$$

$$\tilde{I}_A^* = 20\, e^{-j30}\sigma_1 + 10\, e^{j60}\sigma_2 + 10\, e^{-j60}\sigma_3$$

(30)

Active, reactive, and distortion powers components can now be obtained from (26) as

$$H_A = \begin{pmatrix} 3464.1 - j2000 & 1000 + j1732.05 & 1000 - j1732.05 & 0 \\ 4000 & 1732.05 + j1000 & -j2000 & 0 \\ 0 & 0 & 0 & 0 \\ 2000 & j1000 & 866.03 - j500 & 0 \end{pmatrix}$$

This example states that

$\tilde{P}_{1A} = 3464.1\sigma_0$, $\tilde{P}_{2A} = 1732.05\sigma_0$,

$\tilde{Q}_{1A} = -j2000\sigma_0$, $\tilde{Q}_{2A} = j1000\sigma_0$,

$\tilde{D}_{12_A} = (-3000 + j1732)\sigma_{12}$, $\tilde{D}_{13_A} = (1000 - j1732)\sigma_{13}$,

$\tilde{D}_{23_A} = (-j2000)\sigma_{23}$, $\tilde{D}_{41_A} = 2000\sigma_{41}$. $\tilde{D}_{42_A} = (j1000)\sigma_{42}$,

$\tilde{D}_{43_A} = (866.03 - j500)\sigma_{43}$

For $p, q \in \{1, 2\}$, the active and reactive powers $P_{1A}$, $P_{2A}$, $Q_{1A}$, $Q_{2A}$ respectively are associated to fig. 3. The linear distortion component $\tilde{D}_{12_A}$ as well as the nonlinear distortion components $\tilde{D}_{13_A}$, $\tilde{D}_{23_A}$, $\tilde{D}_{41_A}$, $\tilde{D}_{42_A}$ and $\tilde{D}_{43_A}$ with their corresponding direction and sense can be associated to fig.4. The values of $P_A$, $Q_A$, $D_A$ are

$$P_A = 5196.15 \quad Q_A = 1000 \quad D_A = 5099$$

The rms values of voltage and current respectively are given by

$$|\tilde{U}_A|^2 = 200^2 + 200^2 + 100^2 = 9 \cdot 10^4$$

$$|\tilde{I}_A|^2 = 20^2 + 10^2 + 10^2 = 6 \cdot 10^2$$

The values of $P_A^2, Q_A^2, D_A^2$ are found to sum to $|\tilde{S}|^2$, in accordance with eqn. (12)

$$|\tilde{S}_A|^2 = P_A^2 + Q_A^2 + D_A^2 = 54 \cdot 10^6$$

Therefore, apparent volt-amperes $|\tilde{S}_A|$ at the terminals are found from the relation $|\tilde{S}_A|^2 = |\tilde{U}_A|^2 |\tilde{I}_A|^2 = 54 \cdot 10^6$.

If $\alpha_2$ is changed to $\alpha_2 = 0$ in eqn. (28), the instantaneous voltage is modified into

$$u(t)_B = \sqrt{2} \begin{bmatrix} 200\sin\omega t + 200\sin(2\omega t) \\ +100\sin(4\omega t + 30°) \end{bmatrix} \quad (31)$$

In this case, equal load is presumed, and the resulting instantaneous current is given by

$$i(t)_B = \sqrt{2} \begin{bmatrix} 20\sin(\omega t + 30°) + 10\sin(2\omega t - 30°) \\ +10\sin(3\omega t + 60°) \end{bmatrix} \quad (32)$$

Applying (27), **H** matrix is now given by

$$H_B = \begin{pmatrix} 3464.1 - j2000 & 1732.05 + j1000 & 1000 - j1732.05 & 0 \\ 3464.1 - j2000 & 1732.05 + j1000 & 1000 - j1732.05 & 0 \\ 0 & 0 & 0 & 0 \\ 2000 & 500 + j866.03 & 866.03 - j500 & 0 \end{pmatrix}$$

In both cases, the rms values of the voltage and current are respectively $|\tilde{U}_A|^2 = |\tilde{U}_B|^2 = 9 \cdot 10^4$ $|\tilde{I}_A|^2 = |\tilde{I}_B|^2 = 6 \cdot 10^2$. The values of $P, Q, D$ are the same:
$P_A = P_B = 5196.15, \quad Q_A = Q_B = 1000, \quad D_A = D_B = 5099$
and the apparent volt-amperes are

$$|\tilde{S}|_A^2 = |\tilde{S}|_B^2 = U^2 I^2 = 54 \cdot 10^6$$

It should be noted that the new complex-bivector distortion components
$\tilde{D}_{12_B} = (-1732 + j3000)\sigma_{12}$, $\tilde{D}_{23_B} = (1000 - j1732)\sigma_{23}$
$\tilde{D}_{42_B} = (500 + j866.03)\sigma_{42}$
are now different to bivectors $\tilde{D}_{12_A}, \tilde{D}_{23_A}, \tilde{D}_{42_A}$, and do not equal geometrical representation. Consequently, the above components are sensitive to a change of voltage phase-angles, and the waveform distortion affects the distortion power bivector. In other words, the bivector distortion depends on voltage phase-angles, although its magnitude is invariant. These two cases could not be differentiated in terms of Budeanu's equation power, since all their components are identical. Component identification can be made only through the multivectorial apparent power presented. In general, it can be verified that the power multivector concept detects the direction and sense, not only of multivectorial reactive and distortion components but also of active power components. Thus, the possible reverse sense of any harmonic of active power is very important for a correct identification of harmonic source locations [24], [25], and for determining the responsibility of the utility side (generator) and the consumer side (load).

*B. Example 2*

Fig. 5 represents a linear load with a compensator to minimize the nonactive power/current. The compensator is assumed to consist only of passive components (inductors and capacitors). The voltage at point *A* is given by

$$u(t)_A = \sqrt{2} \begin{bmatrix} 200\sin\omega t + 200\sin(2\omega t - 30°) \\ +100\sin(3\omega t + 30°) \end{bmatrix} \quad (33)$$

and the resulting current before compensation is given by

$$i(t)_A = \sqrt{2} \begin{bmatrix} 20\sin(\omega t - 45°) + 10\sin(2\omega t - 60°) \\ +10\sin(3\omega t + 60°) \end{bmatrix} \quad (34)$$

where $\omega = 100\pi \frac{rad}{s}$.

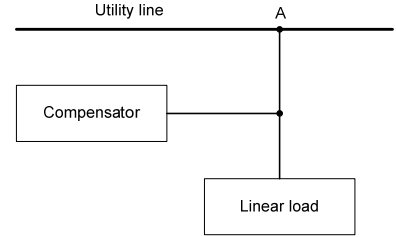

Fig. 5. Shunt compensator configuration

To improve the power factor, let us consider first a shunt capacitor $C_{opt} = 36.53\mu F$ and next a shunt *LC*-branches compensator [26]. In this sense, to obtain a set of parallel branches $L_i$ and $C_i$, the fixed-pole condition is $p_1 = 1.2\omega$, $p_2 = 2.5\omega$ and $p_3 = 4.5\omega$. The simulation results are

$$L_1 = 121mH \quad L_2 = 69mH \quad L_3 = 81mH$$
$$C_1 = 58.29\mu F \quad C_2 = 23.35\mu F \quad C_3 = 6.18\mu F$$

In accordance with (9), and applying (26), apparent power multivector before compensation is given by
$\tilde{S} = (5426.5 + j3328.43)\sigma_0 + (-35.28 - j2131.65)\sigma_{12} +$
$+(-931.85 - j2249.69)\sigma_{13} + (-866.03 - j1500)\sigma_{23}$
and after compensation with *LC*-branches $\tilde{S}$ becomes
$\tilde{S}_{LC} = (5426.5 + j0)\sigma_0 + (-949.51 - j548.2)\sigma_{12} +$
$+(275.22 - j158.9)\sigma_{13} + (433 - j749.98)\sigma_{23}$

A qualitative representation of these power multivectors is depicted in figures (6) and (7).

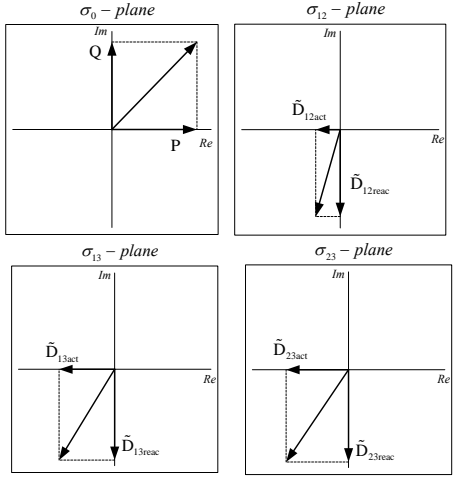

Fig. 6. $\tilde{S}$ components before compensation

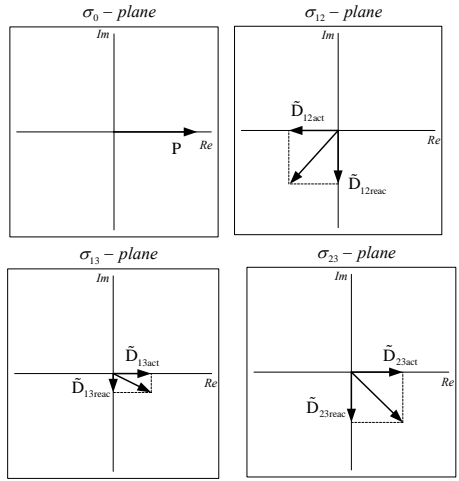

Fig. 7. $\tilde{S}_{LC}$ components after LC compensation

According to (24) and applying (26), multivectorial *relative quality indexes* $(\tilde{\delta})$ and *power factors* (*PF*) are achieved. First, they are calculated prior to compensation $(\tilde{\delta}, PF)$, second for an optimal capacitive (C$_{opt}$) compensator, $(\tilde{\delta}_{Copt}, PF_{Copt})$, and finally for *LC*-branches compensator $(\tilde{\delta}_{LC}, PF_{LC})$.

$\tilde{\delta} = (1+j0.61)\sigma_0 + (-0.007-j0.39)\sigma_{12} + (-0.17-j0.42)\sigma_{13} +$
$+(-0.16-j0.28)\sigma_{23} \Rightarrow |\tilde{\delta}| = 1.35 \Rightarrow PF = 0.74$

$\tilde{\delta}_{Copt} = (1+j0.30)\sigma_0 + (0.04-j0.47)\sigma_{12} + (-0.21-j0.49)\sigma_{13} +$
$+(-0.20-j0.30)\sigma_{23} \Rightarrow |\tilde{\delta}_{Copt}| = 1.31 \Rightarrow PF_{Copt} = 0.76$

$\tilde{\delta}_{LC} = (1+j0)\sigma_0 + (-0.18-j0.10)\sigma_{12} + (0.05-j0.03)\sigma_{13} +$
$+(0.08-j0.14)\sigma_{23} \Rightarrow |\tilde{\delta}_{LC}| = 1.03 \Rightarrow PF_{LC} = 0.97$

TABLE I

|  | - | $C_{opt}$ | LC |
|---|---|---|---|
| $I_{Load}$ | 24.5 | 23.702 | 18.708 |
| P | 5426.50 | 5426.5 | 5426 |
| Q | 3328.43 | 1606.83 | 0 |
| D12 | 2131.94 | 2536.73 | 1096.36 |
| D13 | 2435.05 | 2890.84 | 316.23 |
| D23 | 1732.05 | 1934.25 | 866.02 |
| S | 7348.47 | 7110.69 | 5612 |
| $|\tilde{\delta}|$ | 1.35 | 1.31 | 1.03 |
| PF | 0.74 | 0.76 | 0.97 |

The compensation results listed in Table I show that the *LC* compensator is much more effective in each of the scalar and bivector components than is capacitive compensation (*C$_{opt}$*). Optimal compensation of nonactive power could require suitable power decompositions. In this sense, the new suggested multivectorial apparent power $\tilde{S}$ possesses clear advantages from the viewpoint of nonactive minimization. The principal advantage is that $\tilde{S}$ is decomposed into complex-scalar and complex-bivectors with direction and sense. These components provide detailed information for minimization of each power term by means of new devices, strategies, and algorithms. The accomplishment of such compensating methods and devices is a problem that warrants further research.

## VII. CONCLUSIONS

In this paper, a new multivectorial representation to the apparent power under periodic *n*-sinusoidal/ linear- nonlinear operation has been presented. It is based on an appropriate mathematical tool *(Generalized Complex Geometric Algebra)* to define the power multivector concept in frequency domain. In this mathematical environment, the apparent power multivector $\tilde{S}$ is defined without arbitrariness, it is uniquely determined in a natural way from a new generalized geometric product of voltage and conjugated current vector-phasors. This original quantiy condenses all the information needed to solve future problems on power theory and plays a similar role to the Steinmetz phasor model in sinusoidal case.

The proposed representation is unified and internally consistent with existing power equations. The application of these new concepts to the electric circuit theory should make significant improvements possible in compensating devices, new optimization algorithms and effective power quality indexes.

At last, the opportunity of extending power multivector concept in polyphase systems is entirely possible.

## APPENDIX: GENERALIZED COMPLEX GEOMETRIC ALGEBRA

### A. Generalized complex geometric product: Introduction

We define as $C$ the complex-vector space and $Cl_n$ the Clifford algebra on $n$-dimensional real space $V^n$. We define the set

$$CCl_n = \left\{ \sum_{k=1,2...n} \overline{z}_{1...k} \sigma_{1...k} \right\} \quad (A1)$$

where the coefficients $\overline{z}_{1...k} \in C$ and the basis $\sigma_{1...k} \in Cl_n$. It is trivial that $CCl_n$ is a vector space over $R$. According to (A1) definition, in complex-vectors case we obtain the vector subspace $[CCl_n]_1 = \sum_{p=1}^{n} \overline{z}_p \sigma_p$, where $\overline{z}_p \in C$ and $\sigma_p \in Cl_n$. The generic element $\overline{z}_p \sigma_p$, is a $p$-th complex-vector, and it can be represented by $(a_p + jb_p)\sigma_p$. In complex-bivectors case, we obtain the vector subspace $[CCl_n]_2 = \sum_{p \neq q} \overline{z}_{pq} \sigma_{pq}$. The generic element $\overline{z}_{pq} \sigma_{pq}$, is a $pq$-th complex-bivector, and it can be represented by $(a_{pq} + jb_{pq})\sigma_{pq}$. In the most general form, complex-multivectors, we obtain the vector subspace $[CCl_n]_k = \sum \overline{z}_{12...k} \sigma_{12...k}$. The element $\overline{z}_{12...k} \sigma_{12...k}$, is the $12...k$-th complex-multivector, and it may be represented by $(a_{12...k} + jb_{12...k})\sigma_{12...k}$. Therefore, $CCl_n$ equation (A1), also can be represented as

$$CCl_n = \underbrace{C}_{\substack{complex \\ scalar}} \oplus \underbrace{[CCl_n]_1}_{\substack{complex \\ vectors}} \oplus \underbrace{[CCl_n]_2}_{\substack{complex \\ bivectors}} \oplus \cdots \oplus \underbrace{[CCl_n]_n}_{\substack{complex \\ pseudoscalar}}$$

The structure $\{CCl_n, \hat{g}\}$ is a complex geometric algebra because the following properties are fulfilled: associative, distributive with respect to the sum and contraction.

### B. Generalized complex geometric product for vectors.

Let $\{\sigma_1,...,\sigma_n\}$ a vector basis of $CCl_n$. For two vectors $\tilde{Z}_p = \overline{z}_p \sigma_p$ $(p \in \Omega)$ and $\tilde{Z}'_q = \overline{z}'_q \sigma_q$ $(q \in \Psi)$ where $\Omega, \Psi \subseteq \{1,2,...,n\}$, and complex numbers associated to each vector are

$$\overline{z}_p = |\overline{z}_p| e^{j\alpha_p}$$

$$\overline{z}'_q = |\overline{z}'_q| e^{j\beta_q} = |\overline{z}'_q| e^{j(\alpha_q - \varphi_q)} \quad (B1)$$

we define a new geometric product termed *"generalized complex geometric product"*, $\hat{g}$:

$$\hat{g} : \left( \Re_{\alpha_p, \alpha_q} \circ g \right) \quad (B2)$$

The letter "$g$" represents the usual geometric product and $\Re_{\alpha_p, \alpha_q}$ is an application in the complex planes associated to any multivector product when $\alpha_p \neq \alpha_q$ and it is given by

$$\Re_{\alpha_p, \alpha_q} (\overline{z}_p, \overline{z}'_q) = \begin{cases} e^{-2j(\alpha_q - \alpha_p)} & \text{if } p > q, \; p, q \in N \\ 1 \text{ otherwise}, & p \text{ and/or } q \notin N \end{cases} \quad (B3)$$

where $N = \Omega \cap \Psi$.

This new product for vectors $\tilde{Z}_p$ and $\tilde{Z}_q$ is given by

$$\overline{z}_p \sigma_p \hat{g} \overline{z}'_q \sigma_q = \overline{z}_p \overline{z}'_q \sigma_{pq} \quad (B4)$$

and the basis transposition holds that

$$\left( \overline{z}'_q \overline{z}_p \sigma_{qp} \right) = (-1) \Re_{\alpha_p, \alpha_q} \overline{z}_p \overline{z}'_q \sigma_{pq} \quad (B5)$$

Note that the transposition operation is involutive.
If $\alpha_p = \alpha_q \;\; \forall \; p, q \in N$, then

$$\Re_{\alpha_p, \alpha_p} = Id_C \quad (B6)$$

and "$\hat{g}$", (B2), will then become the classic geometric product "$g$" (3). It should be noted that when $C$ is restricted to real numbers, the classic Clifford Algebra is obtained.

In particular, for two complex-vectors

$$\tilde{Z} = \sum_p |\overline{z}_p| e^{j\alpha_p} \sigma_p \text{ and } \tilde{Z}' = \sum_q |\overline{z}'_q| e^{j(-\alpha_q + \varphi_q)} \sigma_q,$$

where the angles $\alpha_p$ and $(-\alpha_q + \varphi_q)$ identify the phase of the $p$-th and $q$-th harmonics respectively, the generalized complex geometric product in linear operation ($p, q \in N$), can be written

$$\tilde{Z} \hat{g} \tilde{Z}' = \sum_p |\overline{z}_p||\overline{z}'_p| e^{j\varphi_p} + \sum_{p<q} e^{j(\alpha_p - \alpha_q)} |\overline{z}_p||\overline{z}'_q| e^{j\varphi_q} \sigma_{pq} +$$
$$+ \sum_{q<p} e^{j(\alpha_q - \alpha_p)} |\overline{z}_q||\overline{z}'_p| e^{j\varphi_p} \sigma_{qp} = \sum_p |\overline{z}_p||\overline{z}'_p| e^{j\varphi_p} + \quad (B7)$$
$$+ \sum_{p<q} \left\{ e^{j(\alpha_p - \alpha_q)} |\overline{z}_p||\overline{z}'_q| e^{j\varphi_q} - \Re_{\alpha_p, \alpha_q} e^{j(\alpha_q - \alpha_p)} |\overline{z}_q||\overline{z}'_p| e^{j\varphi_p} \right\} \sigma_{pq}$$

where

$$\Re_{\alpha_p, \alpha_q} e^{j(\alpha_q - \alpha_p)} |\overline{z}_q||\overline{z}'_p| e^{j\varphi_p} \sigma_{qp} = e^{j(\alpha_p - \alpha_q)} |\overline{z}_q||\overline{z}'_p| e^{j\varphi_p} \sigma_{qp}$$

### C. Reverse and conjugated operations

We define the bivector *reverse* element as

$$\left( \overline{z}_{qp} \sigma_{qp} \right)^{\dagger} = (-1) \overline{z}_{pq} \sigma_{pq} \quad (C1)$$

where ($\dagger$) is the "*reverse*" operation.

The "*conjugated*" operation ($*$) is given by

$$\left( \overline{z}_p \sigma_p \right)^* = \overline{z}^* \sigma_p \quad (C2)$$

### D. Norm definition.

The *norm, value or magnitude*, of a multivector $\tilde{Z}$ is a unique scalar $\|\tilde{Z}\|$ calculated by

$$\|\tilde{Z}\|^2 = \langle \tilde{Z}(\tilde{Z}^\dagger)^* \rangle_0 \qquad (D1)$$

where we apply $(*)$ in $C$, and $(\dagger)$ in $Cl_n$.

### E. Clifford Fourier transform

Let a continue signal $f_k : R \to C$, $f_k(t) = X_k e^{j(\omega_k t + \theta_k)}$. The Fourier transform $(\Gamma_F)$ of $f_k$ is given by

$$\Gamma_F \{f_k(t)\}(\omega) = \frac{1}{T} \int_T f_k(t) e^{-j\omega_k t} dt = X_k e^{j\theta_k} \qquad (E1)$$

where $j^2 = -1$.

Let $\tilde{f} : R \to CCl_n$ a real-valued multivectorial function. Then $\tilde{f}(t) = \sum_{A \in P(\{1,...,n\}) \cup 0} f_A(t) \sigma_A$ with $f_A(t) = X_A e^{j(\omega_A t + \theta_A)}$, where $P(\{1,...,n\})$ is the set of the all subsets of $\{1,...,n\}$.

According to the linearity of the Clifford Fourier transform $(\Gamma_{C-F})$, we get:

$$\Gamma_{C-F} \{\tilde{f}(t)\}(\omega) = \sum_{A \in P(\{1,...,n\}) \cup 0} \Gamma_F \{f_A(t)\} \sigma_A \qquad (E2)$$

and

$$\Gamma_{C-F} \{\tilde{f}(t)\}(\omega) = \sum_{A \in P(\{1,...,n\}) \cup 0} X_A e^{j\theta_A} \sigma_A = \tilde{F}(\omega) \qquad (E3)$$

where $\tilde{F}(\omega)$ is a *multivector phasor*.

### F. Directed Fourier series associated to voltage and current elements.

We define directed Fourier series associated to $u(t)$ and $i(t)$ voltage and current nonsinusoidal functions as

$$\tilde{U}(t) = \sum_p U_p e^{j(\omega_p t + \alpha_p)} \sigma_p$$

$$\tilde{I}(t) = \sum_q I_q e^{j(\omega_q t + \alpha_q - \phi_q)} \sigma_q \qquad (F1)$$

respectively. Applying the Clifford Fourier transform (E2), we obtain

$$\tilde{U} = \sum_p U_p e^{j\alpha_p} \sigma_p \text{ and } \tilde{I} = \sum_q I_q e^{j(\alpha_q - \phi_q)} \sigma_q \qquad (F2)$$

where $\tilde{U}$ and $\tilde{I}$ are the voltage and current vector-phasors in $CCl_n$ respectively.

### G. Squared value of apparent power

$$|\tilde{S}|^2 = |\tilde{U} \hat{g} \tilde{I}^*|^2 =$$

$$= \sum_{p,q} \left|(\tilde{U}_p \hat{g} \tilde{I}_p^* + \tilde{U}_q \hat{g} \tilde{I}_q^*)\right|^2 + \sum_{p \neq q} \left|(\tilde{U}_p \hat{g} \tilde{I}_q^* - \tilde{U}_q \hat{g} \tilde{I}_p^*)\right|^2 =$$

$$= \sum_{p \in N \cup L} U_p^2 \sum_{q \in N \cup M} I_q^2$$

(G1)

### List of symbols

| | |
|---|---|
| $n$-sinusoidal | = non sinusoidal or multi-sinusoidal. |
| $V^n$ | = linear space over real numbers |
| $\sigma_{1...k}$ | = basis of the Clifford algebra |
| $\bar{z}_{1...k}$ | = complex numbers $\in C$ or Clifford's coefficients. |
| $\tilde{A}_0$ | = complex-scalar |
| $\tilde{Z}_p$ | = harmonic vector-phasor (complex-vector) |
| $\tilde{Z}_{pq}$ | = complex-bivector |
| $\tilde{Z}$ | = multivector |
| $\Re(\alpha_p, \alpha_q)$ | = (associated rotation to multivectorial planes) |
| $\hat{g}$ | = geometric product with associated rotation |
| $|\tilde{Z}_p|, Z_p$ | = magnitude of *p*-th multivector-phasor |
| $\|\tilde{Z}\|$ | = norm |
| $(\tilde{Z})^\dagger$ | $= (-1)^{k(k-1)/2} \tilde{C}$ (reverse element) |
| $a \cdot b$ | = inner product |
| $a \wedge b$ | = outer product or bivector |
| $\tilde{U}_p$ | = *p*-th harmonic voltage vector-phasor |
| $\tilde{I}_p$ | = *p*-th harmonic current vector-phasor |
| $\tilde{S}$ | = multivectorial apparent power |
| $\tilde{D}_{Lin}$ | = linear distortion power bivector |
| $\tilde{D}_{Nonlin}$ | = nonlinear distortion power bivector |
| $\tilde{D}$ | = distortion power bivector |
| $|\tilde{D}|, D$ | = magnitude of distortion power bivector |
| $\sigma_0$ plane | = scalar plane |
| $\sigma_{pq}$ plane | = bivectorial plane |
| $R$ | = real numbers |
| $C$ | = complex vector space |
| $Cl_n$ | = Clifford algebra in *n*-dimensional real space |
| $\alpha_p$ | = phase angle of *p*-th voltage vector-phasor |
| $\alpha_q$ | = phase angle of *q*-th current vector-phasor |
| $\varphi_q$ | = phase angle between q-th voltage vector-phasor angle and q-th current vector-phasor angle. |
| * | = conjugated operation |
| $\Gamma_F$ | = Fourier transform |
| $\Gamma_{C-F}$ | = Clifford Fourier transform |
| H | = complex coefficients matrix for representing the geometric product. |
| $\tilde{\delta}$ | = relative quality index multivector (*RQI*) |
| PF | = power factor |

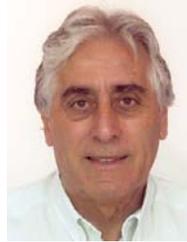

**M. Castilla** ((M'90, M´00) was born in Albaida (Sevilla-Spain). He received the PhD. degree in Physical Sciences from the University of Seville, Spain, in 1989. He is currently Professor of Circuits Theory and Power System Harmonics with the University of Sevilla, Spain.

He has participated in different projects related to Power Quality, and the control of Industrial processes. His research interests have been devote to analysis and electrical measurement in power systems in *n*-sinusoidal conditions, power transfer quality, assessment via wavelet transform analysis and Mathematical tools for analysis power theory. He is a member of the Invespot Group on Electrical and Electronic Measurements.

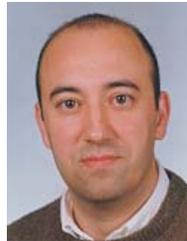

**Juan Carlos Bravo** received the "Licenciatura" in Physical Sciences from the University of Seville, Spain, in 1998, where he is currently Professor of electrical engineering and working towards the Ph.D. degree.

His research interests include signal processing and signal analysis via time-frequency transforms and geometric algebra applied to power quality in polyphase systems.

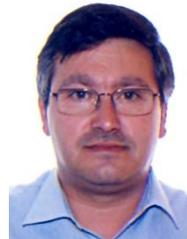

**M. Ordoñez** received the PhD degree in Mathematical Sciences from the University of Seville, Spain, in 1996. He has been Professor in the Applied Mathematics Department.

His research interests include Lie Algebras, Clifford Algebras and applications, and Game Theory.

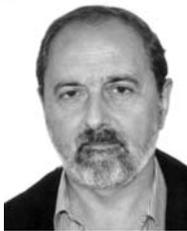

**Juan Carlos Montaño** (M'80, SM'00) received the PhD degree in physics from the University of Seville, Spain, in 1972.

From 1973 to 1978 he was a Researcher at the Instituto de Automática Industrial (CSIC - Spanish Research Council), Madrid, Spain, working on analog signal processing, electrical measurements, and control of industrial processes. Since 1978, he has been responsible for various projects in connection with research in power theory of nonsinusoidal systems, reactive power control, and power quality at the Spanish Research Council.